\documentclass{article}
\usepackage{ijcai26}

\usepackage{times}
\usepackage{soul}
\usepackage{url}
\usepackage[hidelinks]{hyperref}
\usepackage[utf8]{inputenc}
\usepackage[small]{caption}
\usepackage{graphicx}
\usepackage{amsmath}
\usepackage{amsthm}
\usepackage{booktabs}
\usepackage{algorithm}
\usepackage{algorithmic}
\usepackage{amsmath}
\usepackage{amsthm}
\usepackage{amssymb} % provides \mathbb
\usepackage{colortbl}
\usepackage{xcolor}
\usepackage[table]{xcolor}
\usepackage{makecell}
\usepackage{multirow}
\usepackage{float}

\usepackage[switch]{lineno}

\title{Towards Explicit Acoustic Evidence Perception in Audio LLMs for Speech Deepfake Detection}

\author{
Xiaoxuan Guo$^{1,2}$\thanks{These authors contributed equally to this work.}
\and
Yuankun Xie$^{1,2,*}$
\and
Haonan Cheng$^1$
\and
Jiayi Zhou$^2$
\and
Jian Liu$^2$
\and \\
Hengyan Huang$^1$
\and
Long Ye$^3$
\and
Qin Zhang$^3$ \\
\affiliations
$^1$ State Key Laboratory of Media Convergence and Communication, Communication University of China\\
$^2$ Machine Intelligence, Ant Group\\
$^3$ Key Laboratory of Media Audio \& Video, Ministry of Education, Communication University of China\\
\emails{xiaoxuanguo@mails.cuc.edu.cn}
}

\begin{document}

\maketitle

\begin{abstract}
Speech deepfake detection (SDD) focuses on identifying whether a given speech signal is genuine or has been synthetically generated. Existing audio large language model (LLM)-based methods excel in content understanding; however, their predictions are often biased toward semantically correlated cues, which results in fine-grained acoustic artifacts being overlooked during the decision-making process. Consequently, fake speech with natural semantics can bypass detectors despite harboring subtle acoustic anomalies; this suggests that the challenge stems not from the absence of acoustic data, but from its inadequate accessibility when semantic-dominant reasoning prevails. To address this issue, we investigate SDD within the audio LLM paradigm and introduce SDD with Auditory Perception–enhanced Audio Large Language Model (SDD-APALLM), an acoustically enhanced framework designed to explicitly expose fine-grained time--frequency evidence as accessible acoustic cues. By combining raw audio with structured spectrograms, the proposed framework empowers audio LLMs to more effectively capture subtle acoustic inconsistencies without compromising their semantic understanding. Experimental results indicate consistent gains in detection accuracy and robustness, especially in cases where semantic cues are misleading. Further analysis reveals that these improvements stem from a coordinated utilization of semantic and acoustic information, as opposed to simple modality aggregation. 
\end{abstract}

\begin{figure}[t] %强制图片放置于页面顶部
	\centering
	\includegraphics[width=0.94\linewidth]{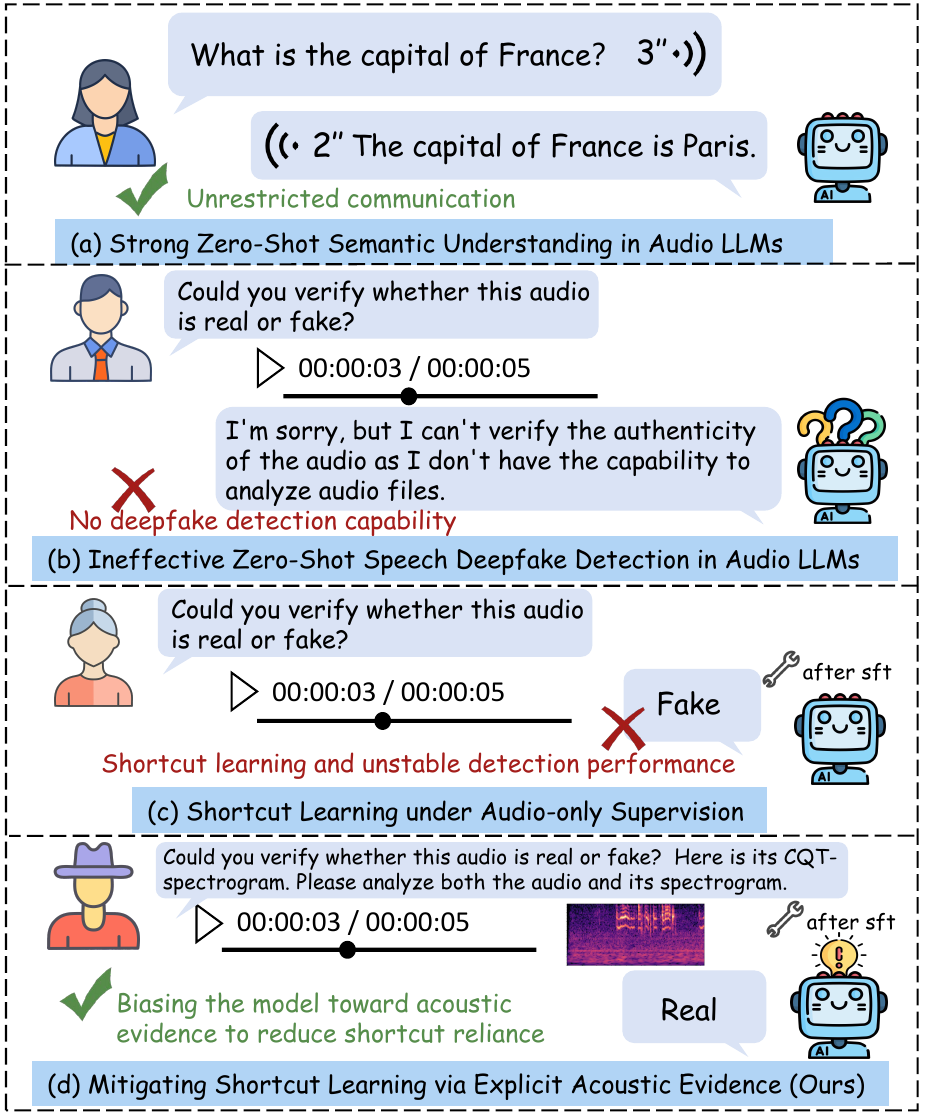}
	\caption{Illustration of the capability gap of audio LLMs in speech deepfake detection.
While audio LLMs exhibit strong semantic understanding, they struggle with reliable deepfake detection when acoustic evidence is accessed implicitly.
Introducing explicit time--frequency representations reshapes acoustic evidence access, leading to more stable and reliable detection.}
        \vspace{-10pt}
	\label{fig-juanshou}
\end{figure}

\section{Introduction}

The task of Speech deepfake detection (SDD) is to ascertain if a speech signal is genuine or spoofed, given that spoofed audio is derived from speech synthesis or voice conversion. With the rapid advancement of speech generation technologies \cite{zhang2025preference,gao2025ttslow,saeki2023zerospeech}, highly realistic synthetic speech can be easily produced, making reliable SDD essential for the security and trustworthiness of speech-based applications.

Most existing SDD techniques prioritize modeling the acoustic characteristics of speech signals; they leverage handcrafted features~\cite{xie2023domain,pham2024deepfake} or learned representations~\cite{tak2022wav2vec,martin2022vicomtech,zhang2024xlsr} to capture artifacts resulting from speech synthesis~\cite{li2025dialogueagents} and voice conversion~\cite{yao2025stablevc}.  Recent research has explored audio large language models (LLMs) for SDD, benefiting from their powerful representation learning capacity and streamlined end-to-end processing. However, since these models are pretrained with semantic- and instruction-oriented objectives, they focus on speech understanding and cross-modal alignment instead of forensic discrimination. Consequently, despite acoustic information being implicitly encoded~\cite{li2025dfallm}, fine-grained spoof-related artifacts are frequently difficult to access during inference, especially when the spoofed speech is semantically natural. Furthermore, audio-only supervised fine-tuning (SFT) fails to provide a fundamental solution to this constraint; rather, it may encourage a dependence on spurious cues, which results in inconsistent detection performance. This issue is depicted in Figure~\ref{fig-juanshou}.

To address this limitation, we re-examine SDD within the audio LLM paradigm through the lens of \emph{acoustic evidence accessibility}. We contend that the primary challenge lies not in the lack of acoustic information, but in the restricted access to discriminative acoustic evidence throughout the model reasoning process. In conventional audio-only SFT, audio LLMs are inclined to rely on semantically dominant representations, which results in shortcut learning and unstable detection behavior. This necessitates mechanisms that explicitly steer audio LLMs toward fine-grained acoustic evidence during inference, instead of depending exclusively on raw waveform inputs.

Guided by this perspective, we introduce \textbf{SDD with Auditory Perception–enhanced Audio Large Language Model (SDD-APALLM)}, an acoustically enhanced framework designed for speech deepfake detection within the audio LLM paradigm. Without introducing additional information sources or altering the pretrained audio encoder, \textbf{SDD-APALLM} enhances \emph{acoustic evidence accessibility} by explicitly providing structured time--frequency representations alongside raw audio, allowing fine-grained acoustic cues to be more easily perceived and exploited during inference. Specifically, we adopt Constant-Q Transform (CQT)–based visual tokens as an explicit evidence representation, as its logarithmic frequency resolution aligns with pitch perception and highlights localized harmonic structures that are closely related to speech synthesis artifacts. By jointly reasoning over raw audio and these explicit cues, the model shifts focus from semantic plausibility toward acoustically grounded evidence, bolstering robustness against semantically natural fakes.

Our main contributions are summarized as follows:

\begin{itemize}
    \item We identify a critical limitation in audio LLM-based speech deepfake detection, observing that audio-only supervised fine-tuning tends to induce a reliance on semantically correlated shortcut cues, this results in strong in-domain performance but unstable generalization across domain shifts.

    \item We introduce SDD-APALLM, an acoustically enhanced detection framework designed to explicitly present time--frequency acoustic evidence to audio LLMs, thereby steering model reasoning toward fine-grained acoustic cues and away from semantic-dominant shortcuts.

    \item Extensive in-domain and cross-domain experiments reveal that enhancing the accessibility of acoustic evidence effectively mitigates performance degradation under domain shift and achieves more stable and robust detection compared to audio-only supervision.
\end{itemize}

\section{Related Work}

\subsection{Acoustic-Centric Speech Deepfake Detection}

Acoustic-centric approaches detect deepfakes based on the assumption that signal-level artifacts are independent of semantic content. Early countermeasures integrated LCNNs with handcrafted features \cite{lavrentyeva2019stc} and CQCCs \cite{todisco2017cqcc} to exploit variable spectro-temporal resolutions. Recent techniques have moved toward end-to-end modeling, such as AASIST \cite{jung2022aasist}, which applies graph attention to raw waveforms, and self-supervised wav2vec~2.0 frameworks \cite{baevski2020wav2vec,tak2022wav2vec} designed for robustness against unseen attacks. Despite performing well in-domain, these acoustic-only models frequently struggle with generalization due to the absence of semantic context. This necessitates the development of frameworks that couple rich contextual insights with fine-grained acoustic sensitivity.

\begin{figure*} %强制图片放置于页面顶部
	\centering
	\includegraphics[width=0.95\linewidth]{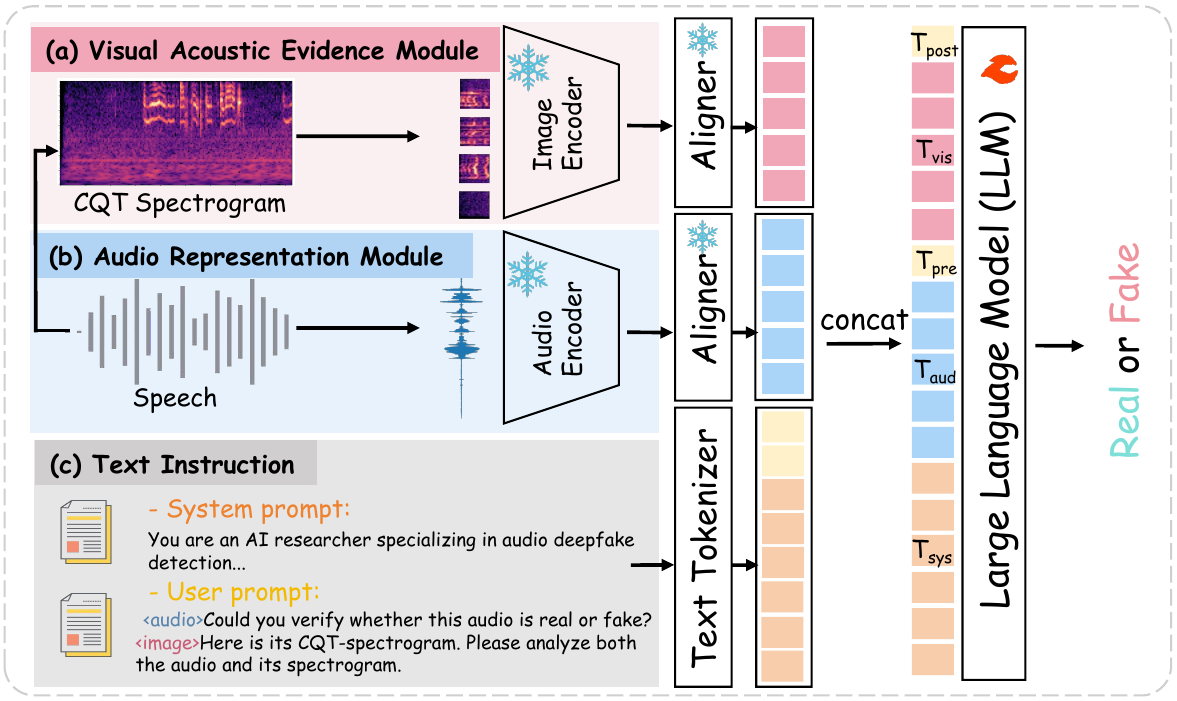}
	\caption{Overview of the proposed SDD-APALLM. The framework combines raw audio and CQT spectrograms to explicitly present fine-grained acoustic evidence through time--frequency representations, facilitating speech deepfake detection within audio LLMs.}
	\label{fig-pipline}
\end{figure*}

\subsection{Multi-View Information Fusion for Speech Deepfake Detection}

Multi-view speech deepfake detection methods seek to enhance robustness by combining heterogeneous speech representations, encompassing acoustic, phonetic, and semantic information. Typical existing methods generate multiple views from the same speech input; these include low-level acoustic features~\cite{todisco2018integrated,yang2019subband,lu2022acoustic}, phoneme-level representations~\cite{zhang2025phoneme,baser2025phonemefake,salvi2025phoneme}, and ASR-based textual embeddings~\cite{wu2024audiomultiview}. Through the alignment and integration of complementary cues from different views, these approaches aim to uncover inconsistencies that might be subtle or view-specific.

Representative works, such as Zhang et al.~\cite{zhang2025phoneme} and Wu et al.~\cite{wu2024audiomultiview}, demonstrate that explicitly integrating phoneme-level artifacts or cross-view correlations—including text and emotion—surpasses purely acoustic models.

However, these frameworks rely on task-specific architectures and explicit fusion, diverging from the Audio LLM paradigm where heterogeneous information is already implicitly embedded. In this context, the challenge shifts from constructing or fusing views to effectively accessing and utilizing fine-grained acoustic evidence during inference.

\subsection{LLM Paradigms for Speech Deepfake Detection}

In contrast to conventional multi-view frameworks that explicitly construct and align multiple representations, audio LLMs implicitly encode acoustic, phonetic, and semantic information within a unified pretrained model. Owing to large-scale pretraining on speech recognition, audio–text alignment, and instruction-following tasks, these models exhibit strong semantic understanding of spoken content and context\cite{ma2025tevatron,liu2025nexus,xie2026interpretable}. However, this implicit multi-view modeling is predominantly shaped by semantic-oriented objectives. Gu et al.~\cite{gu2025allm4add} introduced ALLM4ADD, the first work to apply audio large language models to audio deepfake detection by reformulating detection as an audio question answering task and using supervised fine-tuning to enable robust fake-or-real judgments, especially in data-scarce settings.  Li et al.~\cite{li2025dfallm} improve audio LLM–based deepfake detection by optimizing audio encoder and model components to enhance generalization across multiple detection tasks. Xie et al.~\cite{xie2026interpretable} introduce the first study that provides explicit frequency–time annotation labels for audio deepfake detection, enabling supervised and interpretable training of audio LLMs. As a result, although acoustic information is necessarily involved in downstream decisions, fine-grained acoustic artifacts are often abstracted into high-level representations and become less explicitly accessible during authenticity judgment. This mismatch underscores the necessity of rethinking the way acoustic evidence is presented and utilized in the application of audio LLMs for speech deepfake detection.

\section{Method}

Our method is conceived as a minimal forensic intervention focusing on \emph{acoustic evidence accessibility} within audio LLM reasoning. Instead of incorporating additional information sources or altering the pretrained audio encoder, we explicitly enhance the accessibility of fine-grained acoustic cues to restructure the utilization of acoustic information during inference. By counteracting the dominance of semantic plausibility, stemming from semantic-oriented pretraining and audio-only supervised fine-tuning, the proposed approach steers the model toward attending to acoustically grounded evidence. A comprehensive overview of this framework is depicted in Figure~\ref{fig-pipline}.

\subsection{Speech Representation and Evidence Construction}
To enhance the accessibility of fine-grained acoustic evidence for speech deepfake detection, we represent each utterance through complementary auditory and time--frequency views, both of which are derived from the same source waveform.

\paragraph{Constant-Q Transform.}
Given a discrete waveform $x[n]$, its CQT is computed as
\begin{equation}
X_{\mathrm{CQT}}(f_k, \tau)
= \sum_{n} x[n]\; w_k[n-\tau]\; e^{-j 2\pi f_k n},
\end{equation}
where $f_k$ denotes logarithmically spaced center frequencies with a constant quality factor, $w_k(\cdot)$ is the analysis window, and $\tau$ indexes time frames.
We calculate the CQT magnitude and perform a decibel (dB) transformation, a process designed to amplify the saliency of subtle acoustic artifacts. The resulting map is then normalized and rendered as a pseudo-color image, denoted by $\mathrm{cqt}(x)$.

\paragraph{Sample Construction.}
Each training sample is constructed as
\begin{equation}
\mathcal{S} = \big(x, \mathrm{cqt}(x), y \big),
\end{equation}
where $y \in \{\texttt{real}, \texttt{fake}\}$ indicates bona-fide or spoofed speech.
The waveform inherently retains both semantic and phonetic cues, while the CQT view highlights localized spectral structures and harmonic irregularities associated with speech synthesis and conversion.
Notably, $\mathrm{cqt}(x)$ does not introduce any external information; it is a deterministic re-parameterization of the same signal that makes forensic evidence more accessible to the model.

\subsection{Model Architecture of SDD-APALLM}
We propose SDD-APALLM, an audio LLM--based framework that incorporates CQT evidence to improve sensitivity to subtle acoustic artifacts for speech deepfake detection.
The design bridges the gap between the strong semantic modeling ability of Audio LLMs and the need for explicit access to fine-grained acoustic cues in SDD.
In our implementation, SDD-APALLM is built on Qwen2.5-Omni.
The waveform is processed by the Whisper audio encoder~\cite{radford2023robust}, and the CQT image is processed by a Vision Transformer (ViT).
Both modalities are projected to the shared LLM hidden space via the multimodal aligner and then jointly processed by the LLM.

\paragraph{Audio Representation Module.}
Given a speech waveform $x$, a pretrained audio encoder extracts a sequence of audio tokens
\begin{equation}
\mathbf{T}_{\mathrm{aud}} = f_{\mathrm{aud}}(x) \in \mathbb{R}^{T \times d},
\end{equation}
where $f_{\mathrm{aud}}$ is instantiated as the Whisper audio encoder in Qwen2.5-Omni, producing $T$ audio tokens that are projected to the LLM hidden size $d$.
These tokens encapsulate the semantic, phonetic, and coarse acoustic information acquired through large-scale pretraining. However, as pretraining objectives prioritize semantic content and modality alignment, fine-grained generation artifacts may be attenuated, rendering them less accessible for authenticity judgment.

\paragraph{Visual Acoustic Evidence Module.}
The Visual Acoustic Evidence Module is an evidence-accessibility intervention rather than an independent discriminative pathway.
It reorganizes acoustic cues into a structured time--frequency view that aligns with the multimodal token interface of audio LLMs:
\begin{equation}
\mathbf{T}_{\mathrm{vis}} = f_{\mathrm{vis}}\!\big(\mathrm{cqt}(x)\big) \in \mathbb{R}^{N \times d},
\end{equation}
where $f_{\mathrm{vis}}$ is instantiated as the ViT vision encoder in Qwen2.5-Omni, producing $N$ visual tokens aligned to the same hidden dimension $d$.
The CQT representation retains localized spectral patterns, harmonic structures, and temporal irregularities that are intimately linked to artifacts in speech synthesis and voice conversion. Converting these cues into visual tokens enhances their accessibility within the LLM’s reasoning framework.

\paragraph{Token Integration.}
Let $\mathrm{Tok}(\cdot)$ denote the text tokenizer and $[\cdot;\cdot]$ denote sequence concatenation.
We follow the multimodal prompting template and interleave text prompts with modality tokens.
Specifically, the user prompt is split into two text segments around the \texttt{<image>} placeholder: a prefix $P_{\mathrm{pre}}$ and a post-image instruction $P_{\mathrm{post}}$.
The final input token sequence is constructed as
\begin{equation}
\mathbf{T}(x)=
\Big[
\mathbf{T}_{\mathrm{sys}};
\mathbf{T}_{\mathrm{aud}};
\mathbf{T}_{\mathrm{pre}};
\mathbf{T}_{\mathrm{vis}};
\mathbf{T}_{\mathrm{post}}
\Big],
\end{equation}
where $\mathbf{T}_{\mathrm{sys}}=\mathrm{Tok}(P_{\mathrm{sys}})$, $\mathbf{T}_{\mathrm{pre}}=\mathrm{Tok}(P_{\mathrm{pre}})$, and $\mathbf{T}_{\mathrm{post}}=\mathrm{Tok}(P_{\mathrm{post}})$.
In our setting, $P_{\mathrm{pre}}$ corresponds to the text before the \texttt{<image>} placeholder (e.g., ``Could you verify whether this audio is real or fake?''), and $P_{\mathrm{post}}$ corresponds to the text after it (e.g., ``Here is its CQT-spectrogram. Please analyze both the audio and its spectrogram.'').
This design avoids explicit feature-level fusion and allows the model to integrate multimodal evidence through standard self-attention over a single token sequence. The complete prompt design and examples are provided in Appendix~A.

\paragraph{Training Objective.}
We train the model via SFT with the standard causal language modeling objective.
Given an input audio $x$ and its label $y\in\{\texttt{real},\texttt{fake}\}$, the model is prompted with fixed prompts and learns to generate the label as the completion.
The SFT loss is
\begin{equation}
\mathcal{L}_{\text{SFT}}(\theta)
=
-\mathbb{E}_{(x,y)\sim \mathcal{D}}
\log p_{\theta}\!\left(y \mid \mathbf{T}(x)\right),
\end{equation}
where $p_{\theta}(\cdot)$ denotes the model's next-token distribution conditioned on the input token sequence $\mathbf{T}(x)$.

\begin{table}[]
\centering
\small
\setlength{\tabcolsep}{6pt}
\begin{tabular}{lccc}
\toprule
Model & Training Setting & ACC (\%) & AUC (\%) \\
\midrule
Qwen2.5-Omni-3B & Zero-shot & 9.31 & 59.29 \\
Qwen2.5-Omni-3B & SFT (Audio-only) & \textbf{98.76} & \textbf{99.88} \\
\midrule
Qwen2.5-Omni-7B & Zero-shot & 1.28 & 52.13 \\
Qwen2.5-Omni-7B & SFT (Audio-only) & \textbf{96.07} & \textbf{98.16} \\
\bottomrule
\end{tabular}
\caption{Zero-shot and SFT performance of audio LLMs on ASVspoof2019 LA using audio-only input. Audio LLMs exhibit poor zero-shot capability for speech deepfake detection, while supervised fine-tuning substantially activates discriminative behavior.}
\label{tab:zeroshot_sft}
\end{table}

\newcommand{\Dnineteen}[1]{\textcolor{blue!70!black}{#1}}      % 19LA
\newcommand{\Dntwentyone}[1]{\textcolor{orange!80!black}{#1}}  % 21LA

% Compact "19 / 21" cell formatter
\newcommand{\twods}[2]{\Dnineteen{#1} / \Dntwentyone{#2}}

% Optional: highlight important numbers
\newcommand{\hiN}[1]{\Dnineteen{\textbf{#1}}}   % highlight 19LA
\newcommand{\hiT}[1]{\Dntwentyone{\textbf{#1}}} % highlight 21LA

% add in preamble

\begin{table*}[t]
\centering
\small
\setlength{\tabcolsep}{5pt}
\begin{tabular}{lcccc|c}
\toprule
Input Setting & Acoustic Rep. & ACC (\%)$\uparrow$ & F1 (\%)$\uparrow$ & AUC (\%)$\uparrow$ & Gain (ACC) \\
\midrule

Audio only & -- 
& \twods{98.76}{93.04} 
& \twods{96.80}{74.33} 
& \twods{98.88}{68.51}
& -- \\
\midrule

Acoustic only & \multirow{2}{*}{Mel}
& \twods{91.92}{90.34} 
& \twods{81.83}{64.81} 
& \twods{88.54}{61.55} 
& \multirow{2}{*}{\twods{7.42}{2.51}} \\
\cellcolor{gray!15} Audio + Acoustic & 
& \cellcolor{gray!15} \twods{99.34}{92.85} 
& \cellcolor{gray!15} \twods{98.24}{72.38} 
& \cellcolor{gray!15} \twods{98.95}{66.44} 
&  \\
\midrule

Acoustic only & \multirow{2}{*}{STFT}
& \twods{89.53}{89.05} 
& \twods{79.40}{62.29} 
& \twods{91.50}{60.06} 
& \multirow{2}{*}{\twods{9.41}{4.49}} \\
\cellcolor{gray!15} Audio + Acoustic & 
& \cellcolor{gray!15} \twods{98.94}{\textbf{93.54}} 
& \cellcolor{gray!15} \twods{97.25}{\textbf{76.70}} 
& \cellcolor{gray!15} \twods{98.91}{\textbf{70.65}} 
&  \\
\midrule

Acoustic only & \multirow{2}{*}{LFCC}
& \twods{87.41}{88.91} 
& \twods{73.26}{64.85} 
& \twods{80.15}{62.96} 
& \multirow{2}{*}{\twods{11.46}{4.05}} \\
\cellcolor{gray!15} Audio + Acoustic & 
& \cellcolor{gray!15} \twods{98.87}{92.96} 
& \cellcolor{gray!15} \twods{97.05}{73.49} 
& \cellcolor{gray!15} \twods{98.93}{67.57} 
&  \\
\midrule

Acoustic only & \multirow{2}{*}{MFCC}
& \twods{86.92}{88.54} 
& \twods{70.91}{67.94} 
& \twods{75.76}{67.56} 
& \multirow{2}{*}{\twods{12.27}{4.44}} \\
\cellcolor{gray!15} Audio + Acoustic & 
& \cellcolor{gray!15} \twods{99.19}{92.98} 
& \cellcolor{gray!15} \twods{97.85}{73.33} 
& \cellcolor{gray!15} \twods{98.95}{67.33} 
&  \\
\midrule

Acoustic only & \multirow{2}{*}{CQCC}
& \twods{89.30}{90.53} 
& \twods{75.13}{66.78} 
& \twods{79.38}{63.42} 
& \multirow{2}{*}{\twods{9.79}{2.57}} \\
\cellcolor{gray!15} Audio + Acoustic & 
& \cellcolor{gray!15} \twods{99.09}{93.10} 
& \cellcolor{gray!15} \twods{97.60}{74.08} 
& \cellcolor{gray!15} \twods{\textbf{99.02}}{68.03} 
&  \\
\midrule

Acoustic only & \multirow{2}{*}{CQT}
& \twods{91.49}{91.18} 
& \twods{78.40}{68.69} 
& \twods{80.23}{64.77} 
& \multirow{2}{*}{\twods{7.97}{1.87}} \\
\cellcolor{gray!20} \textbf{Audio + Acoustic} & 
& \cellcolor{gray!20} \twods{\textbf{99.46}}{93.05} 
& \cellcolor{gray!20} \twods{\textbf{98.56}}{73.36} 
& \cellcolor{gray!20} \twods{98.84}{67.23} 
&  \\
\bottomrule
\end{tabular}

\caption{Modality ablation on \Dnineteen{ASVspoof2019 LA} and \Dntwentyone{ASVspoof2021 LA} using Qwen2.5-Omni-3B.
Each cell reports \Dnineteen{ASVspoof2019 LA}/\Dntwentyone{ASVspoof 2021 LA} results.
Gray cells highlight Audio+Acoustic settings, where visualized acoustic representations are introduced as explicit complementary evidence.
Bold numbers indicate the best performance in each column for the corresponding dataset.
The Gain (ACC) column measures the accuracy improvement of Audio+Acoustic over Acoustic-only inputs, i.e., $\Delta=\mathrm{ACC}(\text{Audio+Acoustic})-\mathrm{ACC}(\text{Acoustic-only})$, highlighting how much raw audio amplifies each explicit acoustic representation.}
\label{tab:modality_ablation}
\end{table*}

\section{Experiments}

Beyond overall detection accuracy, our experiments are designed to examine whether explicitly exposing acoustic evidence changes how audio LLMs utilize information during inference. Specifically, we investigate whether supervision alone is sufficient to activate reliable spoof detection, and whether structured time--frequency evidence improves robustness by mitigating over-reliance on semantic plausibility. 

\subsection{Experimental Setup}

\subsubsection{Implementation Details}

We conduct experiments using the Qwen2.5-Omni audio large language models with 3B\footnote{\url{https://huggingface.co/Qwen/Qwen2.5-Omni-3B}} and 7B parameters\footnote{\url{https://huggingface.co/Qwen/Qwen2.5-Omni-7B}}. Models are adapted for speech deepfake detection via SFT with Low-Rank Adaptation (LoRA) applied to all linear layers. During training, the vision encoder and modality aligner are frozen, while the audio LLM backbone remains trainable. This setting allows the model to update its language-level representations while preserving the pretrained audio–visual alignment structure.

Depending on the experimental setting, the model is trained with raw audio only, time--frequency representations only, or joint audio and time--frequency inputs.
Specifically, multiple time--frequency representations, including Mel, STFT, LFCC, MFCC, CQCC, and CQT, are rendered as images and, when applicable, paired with the corresponding audio signals. Task-specific prompts are used to guide binary authenticity classification. Training is conducted using the ms-swift framework \footnote{\url{https://github.com/modelscope/swift}} with distributed training on 8 NVIDIA A100 GPUs, employing bfloat16 precision and gradient checkpointing. The maximum sequence length is set to 768, with per-device batch sizes of 16 for training and 8 for evaluation. The AdamW optimizer is used with a learning rate of $5\times10^{-5}$, weight decay of 0.1, and a warmup ratio of 0.01. All experiments are performed on the ASVspoof2019 LA dataset and evaluated using accuracy (ACC), F1 score, and AUC, with a fixed decision threshold of 0.5 for fair comparison.

\subsubsection{Compared Methods}

To evaluate the performance of our proposed method, we compare it with a diverse set of state-of-the-art speech deepfake detection approaches. These include representative end-to-end audio spoofing models such as RawNet2~\cite{tak2021rawnet2}, RawGAT-ST~\cite{tak2021spectrotemporal}, AASIST and AASIST-L~\cite{jung2022aasist}, TSSDNet~\cite{hua2021towards}, FG~\cite{wang2021investigating},and AMSDF~\cite{wu2024audiomultiview}, which focus on modeling acoustic artifacts using handcrafted or learned representations.

In addition, we include recent audio LLM–based methods, namely ALLM4ADD~\cite{gu2025allm4add} and DFALLM~\cite{li2025dfallm}, which explore the use of large audio–language models for deepfake detection. These baselines are particularly relevant as they share a similar modeling paradigm with our approach. Together, these methods provide a comprehensive benchmark for assessing the effectiveness of explicit acoustic enhancement in Audio LLM–based speech deepfake detection.

\subsection{Zero-Shot and Fine-Tuned Performance of Audio LLMs}

We first examine whether audio LLMs inherently possess speech deepfake detection capability. As shown in Table~\ref{tab:zeroshot_sft}, both Qwen2.5-Omni-3B and 7B perform near random under the zero-shot setting, indicating that strong semantic understanding alone does not translate into reliable spoof detection.

After supervised fine-tuning with audio-only inputs, both models achieve substantial performance gains, demonstrating that task-specific supervision is essential for activating discriminative behavior in audio LLMs. Notably, the 3B model slightly outperforms the 7B variant, suggesting that model scale is not the primary factor once supervision is introduced. These results imply that, while SFT enables audio LLMs to learn spoof-related cues, such cues are acquired implicitly, motivating the need to explicitly expose fine-grained acoustic evidence.

\begin{figure*} %强制图片放置于页面顶部
	\centering
	\includegraphics[width=\linewidth,height=0.42\textheight,keepaspectratio]{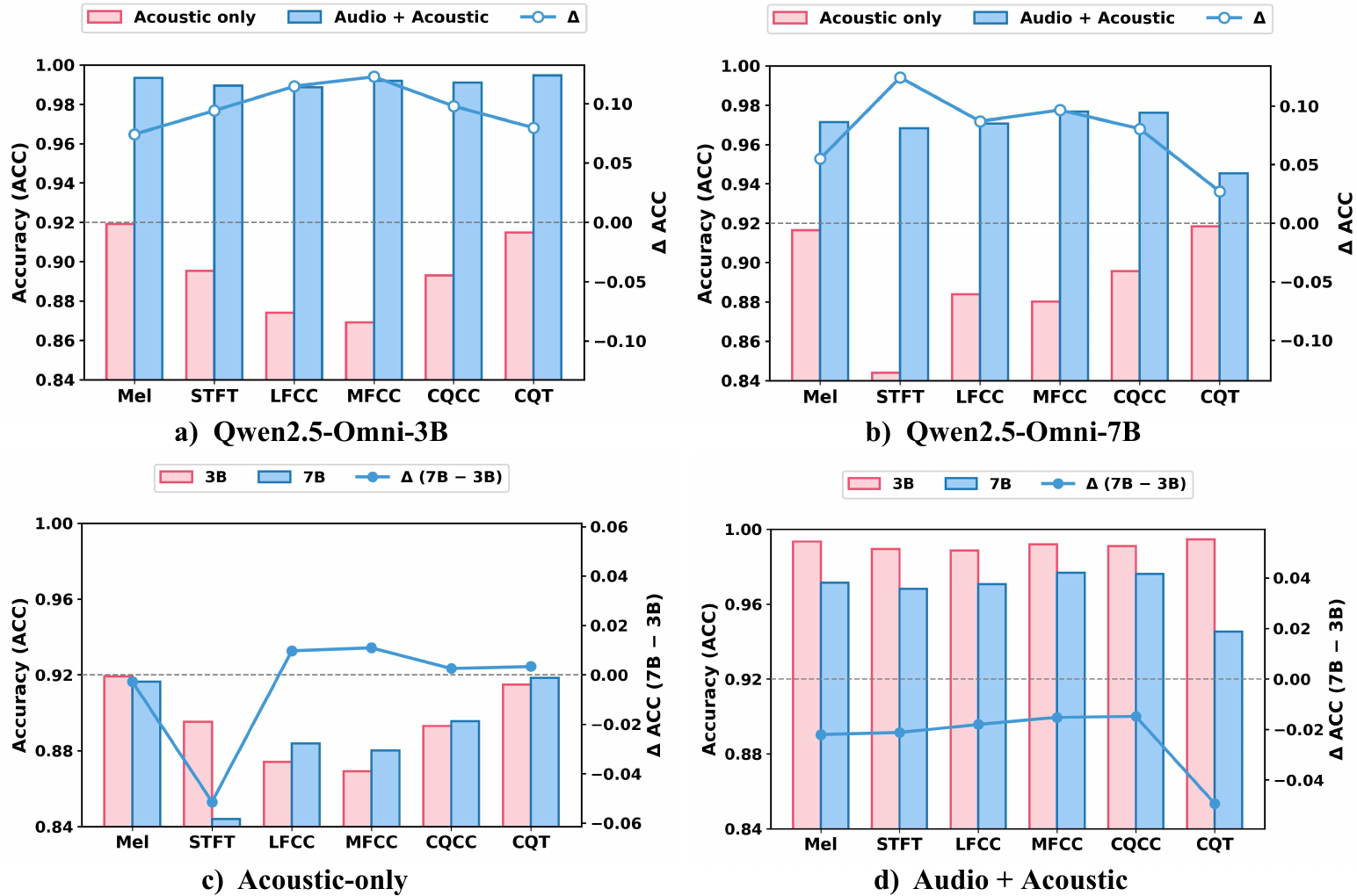}
	\caption{Accuracy analysis across different spectrogram types and model scales.
(a,b) Performance comparison between spectrogram-only and audio--spectrogram inputs for Qwen2.5-Omni-3B and Qwen2.5-Omni-7B, where $\Delta = \mathrm{ACC}(\text{Audio + Spectrogram}) - \mathrm{ACC}(\text{Spectrogram-only})$.
(c,d) Accuracy difference between Qwen2.5-Omni-7B and Qwen2.5-Omni-3B under spectrogram-only and audio--spectrogram settings, computed as $\Delta = \mathrm{ACC}(7\text{B}) - \mathrm{ACC}(3\text{B})$.}
	\label{fig-1}
\end{figure*}

\subsection{Effect of Explicit Acoustic Evidence on audio LLM-based Detection}

We analyze the effect of explicit acoustic evidence via a modality ablation study comparing audio-only, acoustic-only, and audio--acoustic inputs on ASVspoof2019 LA~\cite{todisco2019asvspoof} and ASVspoof 2021 LA~\cite{yamagishi2021asvspoof}.

On the in-domain dataset (ASVspoof2019 LA), acoustic-only inputs perform substantially worse than audio-only baselines, while combining acoustic representations with raw audio consistently improves performance. This indicates that explicit acoustic evidence does not act as an independent discriminative modality, but enhances in-domain detection by complementing audio representations learned through supervision.

Under domain shift (ASVspoof2021 LA), however, a contrasting pattern emerges. Audio-only performance degrades noticeably, whereas acoustic-only inputs exhibit significantly smaller accuracy drops, particularly for CQT. At the same time, the gains from audio--acoustic fusion become marginal. This suggests that audio-only supervised fine-tuning encourages reliance on dataset-specific shortcut cues that generalize poorly, while explicit acoustic representations provide more stable and invariant evidence. A more detailed analysis of such shortcut learning behavior is provided in Appendix B.

Importantly, these findings suggest that the benefit of explicit acoustic evidence is not attributable to additional information content, but to a change in how evidence is accessed during inference. Under audio-only supervision, audio LLMs tend to exploit semantically correlated shortcuts that are highly effective in-domain but fragile under distribution shift. By contrast, explicitly exposing structured time--frequency representations constrains the model to attend to localized acoustic patterns that are less entangled with linguistic identity or dataset-specific priors. As a result, the decision process becomes less sensitive to semantic plausibility and more grounded in invariant acoustic cues, yielding improved stability and robustness.

\subsection{Analysis of Acoustic Types and Model Scale}

We further analyze the role of acoustic representations and model scale in audio LLM-based speech deepfake detection. Figures~\ref{fig-1}(a,b) compare acoustic-only and audio--acoustic inputs across different time--frequency representations using Qwen2.5-Omni-3B and 7B.

Across all representations, acoustic-only performance varies, but incorporating raw audio consistently yields positive gains, as reflected by the $\Delta$ values. This indicates that explicit acoustic evidence serves as a complementary grounding signal, rather than an independent discriminative modality, and its benefit is robust to the choice of representation.

While perceptually motivated or high-resolution representations (e.g., CQT and Mel) tend to provide larger gains, no single acoustic representation consistently dominates. This suggests that performance improvements stem primarily from explicitly exposing time--frequency structure, rather than from specific handcrafted feature designs.

Figures~\ref{fig-1}(c,d) further reveal a scale-dependent effect. Under acoustic-only inputs, the 7B model matches or slightly outperforms the 3B variant, indicating that increased capacity can better utilize explicit acoustic cues when shortcut paths are limited. In contrast, under audio--acoustic inputs, the 7B model consistently underperforms the 3B model, suggesting that larger models are more susceptible to amplifying shortcut correlations from raw audio.

These observations further indicate that model scale interacts non-trivially with evidence accessibility. While larger audio LLMs possess greater representational capacity, this capacity can amplify shortcut exploitation when dominant semantic cues remain unconstrained. Explicit acoustic evidence partially counteracts this tendency by anchoring attention to fine-grained spectral structures, thereby regularizing how increased capacity is utilized. This suggests that improving acoustic evidence accessibility is particularly important for large audio LLMs, where semantic shortcut learning is more likely to overshadow forensic cues.

% \begin{figure*} %强制图片放置于页面顶部
% 	\centering
% 	\includegraphics[width=1\linewidth]{figs/2.pdf}
% 	\caption{Accuracy comparison between Qwen2.5-Omni-3B and Qwen2.5-Omni-7B across different spectrogram types using (a) Spectrogram-only inputs and (b) Audio + Spectrogram inputs.
% The curve shows the accuracy difference $\Delta = \mathrm{ACC}(7\text{B}) - \mathrm{ACC}(3\text{B})$.}
% 	\label{fig-2}
% \end{figure*}

\subsection{Comparison with Existing Methods}

\begin{table}[t]
\centering
\small
\setlength{\tabcolsep}{4pt}
\begin{tabular}{lcc}
\toprule
Method & Model Input & ACC(\%)$\uparrow$ \\
\midrule
\multicolumn{3}{l}{\textit{End-to-End Methods}} \\
RawNet2~\cite{tak2021rawnet2} & Audio & 93.12 \\
TSSDNet~\cite{hua2021towards} & Audio & 89.93 \\
RawGAT-ST~\cite{tak2021spectrotemporal} & Audio & 98.14 \\
AASIST~\cite{jung2022aasist} & Audio & 98.03 \\
AASIST-L~\cite{jung2022aasist} & Audio & 97.06 \\
\midrule
\multicolumn{3}{l}{\textit{Conventional Pipeline Methods}} \\
FG~\cite{wang2021investigating} & Hubert & 95.22  \\
AMSDF~\cite{wu2024audiomultiview} & wav2vec2.0 & 96.88 \\
\midrule
\multicolumn{3}{l}{\textit{Audio LLM-based Methods}} \\
ALLM4ADD~\cite{gu2025allm4add} & Audio & 99.39 \\
DFALLM~\cite{li2025dfallm} & Audio & 99.15 \\
\midrule
\multicolumn{3}{l}{\textit{Ours (Audio LLM-based)}} \\
Baseline & Audio & 98.76 \\
\textbf{SDD-APALLM} & \textbf{Audio + CQT} & \textbf{99.46} \\
\bottomrule
\end{tabular}
\caption{Comparison with existing methods on ASVspoof2019 LA (threshold = 0.5). 
All results of our method are obtained using the Qwen2.5-Omni-3B model.}
\label{tab:input_comparison}
\end{table}

% \begin{table}[t]
% \centering
% \small
% \setlength{\tabcolsep}{2pt}
% \begin{tabular}{lccc}
% \toprule
% Method & Model Input & \#Params (M) & ACC(\%)$\uparrow$ \\
% \midrule
% \multicolumn{4}{l}{\textit{Small-scale audio models}} \\
% RawGAT-ST & Audio & 0.44 & 98.14 \\
% AASIST & Audio & 0.30 & 98.03 \\
% AASIST-L & Audio & 0.09 & 97.06 \\
% RawNet2 & Audio & 17.62 & 93.12 \\
% TSSDNet & Audio & 0.35 & 89.93 \\
% AMSDF & Audio + Text + Emotion & 325.41 & 96.88 \\
% \midrule
% \multicolumn{4}{l}{\textit{Audio large language models}} \\
% ALLM4ADD & Audio & $\sim$ 7700 & 99.39 \\
% DFALLM & Audio & $\sim$817 & 99.15 \\
% \midrule
% \multicolumn{4}{l}{\textit{Ours (Audio LLM-based)}} \\
% Ours & Audio & $\sim$ 3000 & 98.76 \\
% \textbf{Ours} & \textbf{Audio + CQT spectrogram} & $\sim$ 3000 & \textbf{99.46} \\
% \bottomrule
% \end{tabular}
% \caption{Comparison with existing methods on ASVspoof2019 LA (threshold = 0.5).
% All results of our method are obtained using the Qwen2.5-Omni-3B model.}
% \label{tab:input_comparison}
% \end{table}

We compare the proposed method with representative speech deepfake detection approaches on ASVspoof2019 LA, including small-scale end-to-end audio models and recent audio LLM-based methods, as summarized in Table~\ref{tab:input_comparison}. Conventional end-to-end models (e.g., RawGAT-ST, AASIST, RawNet2) operate directly on raw waveforms and achieve strong performance by learning task-specific acoustic representations, but they are typically trained from scratch and lack semantic reasoning capability.

By leveraging large-scale pretraining, recent frameworks such as ALLM4ADD~\cite{gu2025allm4add} and DFALLM~\cite{li2025dfallm} have demonstrated competitive performance through an audio-only input modality, indicating that implicit acoustic cues embedded in raw audio can be exploited under sufficient supervision. However, these methods rely on implicit access to acoustic evidence, without explicitly exposing structured time--frequency information.

In contrast, our method explicitly augments audio LLMs with interpretable time--frequency representations, enabling joint reasoning over raw audio and accessible acoustic evidence. Using the same Qwen2.5-Omni-3B backbone, our audio-only setting already achieves competitive performance among audio LLM-based methods, while incorporating CQT spectrograms further improves accuracy to 99.46\%. This result surpasses both small-scale end-to-end models and prior large-model-based approaches, while relying on a lightweight accessibility intervention rather than additional model capacity or task-specific retraining. Moreover, the consistent gains support our claim that explicitly exposing acoustic evidence leads to more stable and reliable detection, rather than merely improving in-domain accuracy.

\subsection{Interpretability}

\begin{figure}[t]
    \centering
    \includegraphics[width=0.9\linewidth]{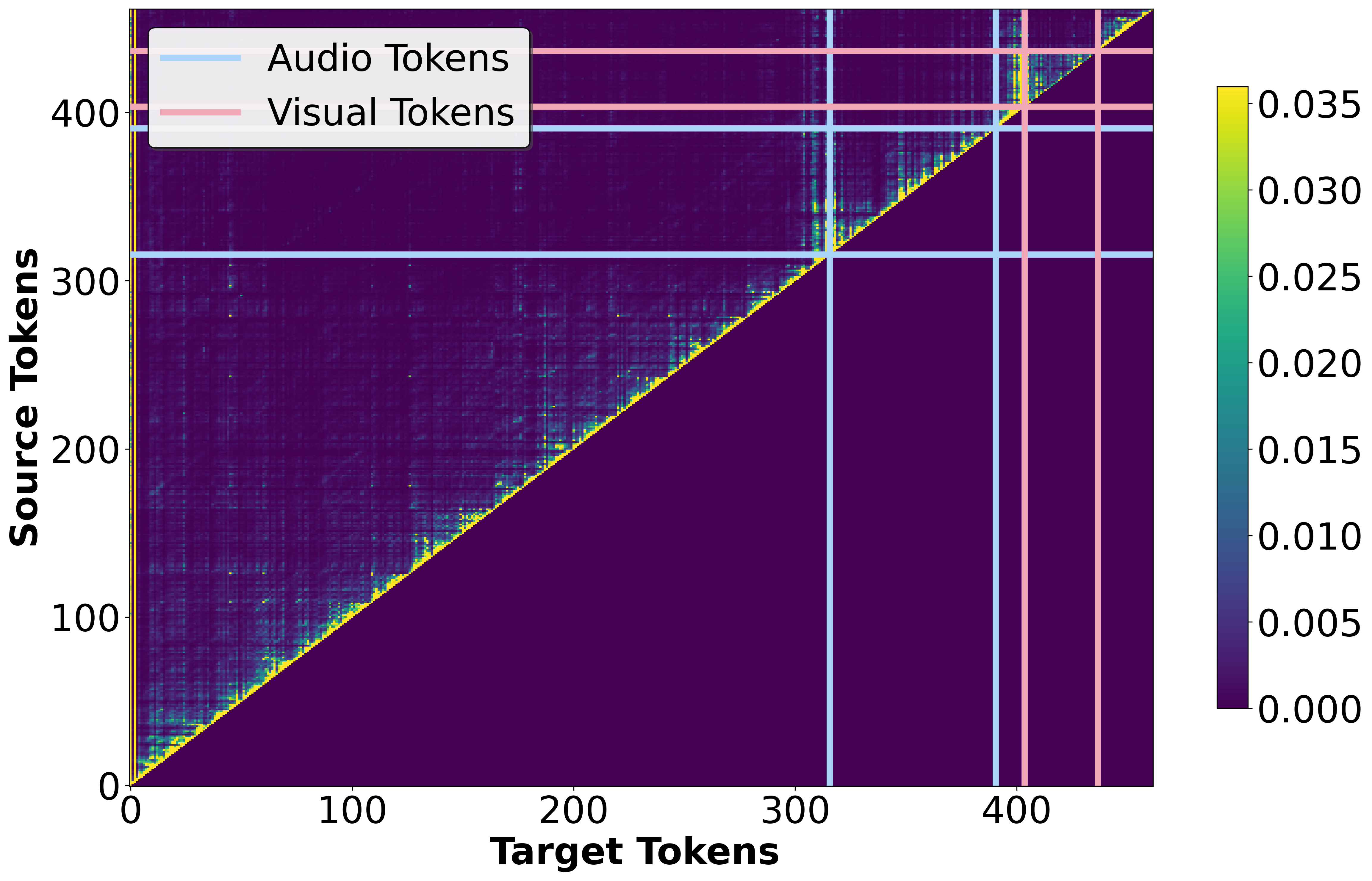}
    \caption{Attention map of SDD-APALLM during inference, illustrating token-level interactions across different input components.
Tokens 1–316 correspond to the system prompt $\mathbf{T}{\mathrm{sys}}$, tokens 317–401 to the audio tokens $\mathbf{T}{\mathrm{aud}}$, tokens 402–414 to the pre-user prompt $\mathbf{T}{\mathrm{pre}}$, tokens 415–450 to the visual tokens $\mathbf{T}{\mathrm{vis}}$, and tokens beyond 450 to the post-user prompt $\mathbf{T}_{\mathrm{post}}$.}
    \label{fig:atten}
\end{figure}

To examine whether SDD-APALLM exploits explicit acoustic evidence during inference, we visualize its attention patterns in Figure~\ref{fig:atten}. The model assigns non-negligible attention not only to the audio-token region but also to tokens corresponding to time--frequency representations, indicating that these acoustic evidence tokens are involved rather than ignored during reasoning.

We further inspect attention distributions across modality-specific token regions. As shown in Figure~\ref{fig:atten}, clear attention mass is allocated to both audio tokens and visual tokens, with the latter exhibiting coherent self-attention patterns and observable interactions with text tokens. This suggests that explicit time--frequency representations function as accessible acoustic evidence integrated into the reasoning process, thereby influencing the final detection decision.

\section{Conclusion}

This work revisits speech deepfake detection with audio LLMs from the perspective of \emph{acoustic evidence accessibility}. We show that the core limitation of existing approaches arises not from missing acoustic information, but from the difficulty of accessing fine-grained acoustic evidence under semantic-dominant training paradigms. By explicitly exposing structured time--frequency representations as accessible acoustic cues, our approach encourages audio LLMs to rely less on semantic plausibility and more on acoustically grounded evidence. Extensive experiments across in-domain, cross-domain, and cross-lingual settings demonstrate that this accessibility-oriented intervention consistently mitigates shortcut learning and reduces performance instability. These results indicate that improving acoustic evidence accessibility offers a simple yet effective strategy for enhancing the robustness, stability, and interpretability of audio LLM-based speech deepfake detection.

% \section*{Ethical Statement}

% There are no ethical issues.

% \section*{Acknowledgments}

\bibliographystyle{named}
\bibliography{ijcai26}

\clearpage
\newpage
\appendix

\end{document}